\documentclass[aps,superscriptaddress,showpacs,nofootinbib]{revtex4}

\usepackage[dvips]{color}                 
\usepackage{eucal}

\usepackage{epsfig}

\begin{document}

\title{Temperature Effects on the Magnetization of Quasi-One-Dimensional\\ Peierls Distorted Materials}

\author{Heron Caldas} \email{hcaldas@ufsj.edu.br} \affiliation{Departamento de
  Ci\^{e}ncias Naturais. Universidade Federal de S\~{a}o Jo\~{a}o del Rei,\\
  36301-160, S\~{a}o Jo\~{a}o del Rei, MG, Brazil}

\begin{abstract}

It is shown that temperature acts to disrupt the magnetization of Peierls distorted quasi-one-dimensional materials (Q1DM). The mean-field finite temperature phase diagram for the field theory model employed is obtained by considering both homogeneous and inhomogeneous condensates. The tricritical points of the second order transition lines of the gap parameter and magnetization are explicitly calculated. It is also shown that in the absence of an external static magnetic field the magnetization is always zero, at any temperature. As expected, temperature does not induce any magnetization effect on Peierls distorted Q1DM.

\end{abstract}


\maketitle

\section{Introduction}

The recent experimental observation of a signature of one-dimensional (1D) transport in 30 nm wide graphene ribbons~\cite{Lin} certainly motivated the theoretical investigation of 1D systems. A very interesting class of Q1DM are the ones where a lattice distortion develops spontaneously with the appearance of a gap $\Delta_0$, rendering the material an insulator. This is the mechanism of the {\it Peierls transition}, which has been observed experimentally in various Q1DM. Probably, one of the most well-known Q1DM that exhibits this phenomenon is {\it trans}-polyacetylene (TPA). This polymer is a 1D chain of $\rm CH$ groups with alternating single and double bonds, having one electron per site. In the tight binding approximation, TPA would be a metal. However, the interaction of the electrons with the lattice (also known as the spontaneous Peierls dimerization) is such that the energy gain in the system is always larger that the energy investment for distorting the lattice. As a consequence, polyacetylene is an insulator. After doping, and the consequent vanishing of the gap at an experimentally measured critical doping concentration $y_c\cong 6 \%$~\cite{Fernando}, the conductivity of TPA is enormously increased, presenting metal-like properties~\cite{Review}. 

Although we shall refer to {\it trans}${-\rm CH}_x$ throughout the paper, due to the large amount of data on this material, the main results obtained here are robust and can be applied to any Q1DM possessing the same dimerized structure of {\it trans}${-\rm CH}_x$ for which a field theory model is suitable to be employed~\cite{Campbell,Zee}. The electron-phonon interaction in {\it trans}${- \rm CH}_x$ is described  by the discrete Su-Schrieffer-Heeger (SSH) Hamiltonian~\cite{SSH}, and its continuum version, the Takayama--Lin-Liu--Maki (TLM) Hamiltonian~\cite{TLM}. The TLM model is a relativistic field theory with two-flavor Dirac fermions that, with a doping equal to or greater than $y_c$, have linear dispersion relations with a Fermi velocity $v_F \approx 10^6 m/s$, which is of the order of the velocity of the Dirac fermions in graphene~\cite{Graphene}. Employing the Gross-Neveu (GN) model~\cite{GN}, that can be properly identified with the TLM model, as an effective field theory model for describing the insulator-metal phase transition in polyacetylene~\cite{CB,CM,PRB}, there has been found a very good agreement with the experimental value, $y_c=\frac{N}{\pi \hbar v_F} a \mu_c \cong 6 \%$~\cite{CM,PRB}, where $N$ (=2 for TPA) is the number of spin degrees of freedom of the (delocalized) $\pi$ electrons, $\hbar$ is the Planck's constant divided by $2\pi$, $v_F=k_F\hbar/m$, $k_F$ is the Fermi wavenumber, $m \equiv \hbar^2/2t_o a^2$, and $a$ ($\cong 1.22 {\rm \AA} $ for TPA) is the lattice (equilibrium) spacing between the $x$ coordinates of successive ${\rm CH}$ radicals in the undimerized structure, and $\mu_c=\frac{\Delta_0}{\sqrt{2}}$ ($\Delta_0$ $\approx 0.7 {\rm eV}$ for TPA) is the critical chemical potential at which the GN model undergoes a first-order phase transition to a symmetry restored (zero gap) phase~\cite{Wolff}. 

Regarding our choice of using a continuum model, a few comments are pertinent here. The TLM model can be derived from the SSH model by expanding its Hamiltonian about the Fermi surface and keeping terms only to lowest order in $a/\xi$, where $\xi \equiv \hbar v_F/\Delta$ is the electronic correlation length. This is fulfilled for the materials we are interested in and will be described by the TLM (GN) model, for which $\xi >> a$. Besides, the clear advantages of employing continuum models such as the TLM (or GN) are the analytical solutions that they provide and the fact that field theory methods are suitable for the calculation of the effective potential, which is appropriate for the analysis of the phase structure of the model, as in this work. On the other hand, the disadvantages are that continuum models with relativistic dispersion relations have an electronic spectrum that is unbounded below, and that the acoustic modes are lost in the continuum limit. The first problem is resolved by adopting a certain energy cutoff, and the second, if only terms to next order in $a/\xi$ are kept~\cite{Review}.

In \cite{NPB1} the zero temperature phase diagram of 1D TPA under asymmetric doping, defined as an imbalance between the chemical potentials of the electrons with the two possible spin orientations (``up" $\equiv \uparrow$, and ``down" $\equiv \downarrow$) introduced in the system by the doping process, has been studied. As emphasized in \cite{NPB1}, the chemical potentials asymmetry between the $\uparrow$ and $\downarrow$ electrons can be achieved experimentally by the actuation of an external static magnetic field $B_0$ on the polyacetylene wire, which breaks the spin-1/2 {\it SU~}(2) symmetry. In \cite{NPB1}, the continuous model that describes the electron-phonon interactions in TPA has been introduced and the magnetization, the critical magnetic field $B_{0,c}$ at which there is a quantum phase transition to a fully polarized (magnetized) phase, and the magnetic susceptibility at zero temperature, within the field theory approach, have been obtained. 

In this paper, we study the thermal effects on the magnetic properties of Peierls distorted Q1DM and verify the possibility of the existence of this fully polarized phase at finite temperature. The mean-field finite temperature phase diagram for the field theory model employed is obtained. The tricritical points of the second-order transition curves of the gap parameter and magnetization are explicitly calculated by considering both homogeneous and inhomogeneous $\Delta(x)$ condensates. One of the main results of this paper is the demonstration of the ``stationarity" of the tricritical point of the second-order transition line of the gap parameter under the the influence of an external (constant) Zeeman magnetic field. In other words, in a Peierls distorted Q1DM under the influence of an external Zeeman magnetic field, the tricritical point obtained considering homogeneous condensates remains at the same location when inhomogeneous $\Delta(x)$ condensates are taken into account.

The paper is organized as follows. In Section II we introduce the model Lagrangian describing polyacetylene. In Section III the temperature dependent renormalized effective potential is presented. In this section we obtain an analytical expression for the effective potential at high temperature, as well as the chemical potentials dependent gap equation and the critical temperature at which the gap vanishes. Besides this, the tricritical points of the second-order transition lines of the gap parameter and magnetization are explicitly calculated. The temperature dependent magnetic properties of Peierls distorted Q1DM are also obtained in this section. In the Summary we present the conclusions.

\section{Model Lagrangian}

For the benefit of the reader, let us reproduce from \cite{NPB1} the model Lagrangian and the basic definitions necessary to describe polyacetylene and equivalent Peierls distorted Q1DM. As we mentioned already, the electron-phonon interaction in TPA is represented by the SSH model~\cite{SSH}, which has a continuum version known as the TLM model \cite{TLM}. The TLM Lagrangian density in the adiabatic approximation (considering static configurations for which $\partial \Delta/\partial t =0$) is given by

\begin{eqnarray}
{\cal L}_{\rm TLM} =
 \sum_{j=1}^N {\psi^j}^\dagger
\left( i \hbar \partial_t - i \hbar v_F \gamma_5 \partial_x - \gamma_0 \Delta(x) 
\right) \psi^j -\frac{1}{2 \pi \hbar v_F \lambda_{\rm TLM}} 
\Delta^2(x)\;,
\label{LagTLM}
\end{eqnarray}
where $\psi$ is a two component Dirac spinor $\psi^j= \left(\begin{array}{cc} \psi^j_{L}\\ \psi^j_{R} \end{array} \right)$, representing the ``left moving" and ``right moving" electrons close to their Fermi energy, respectively, and $j$ is an internal symmetry index (spin) that determines the effective degeneracy of the fermions. We define $1=\uparrow$, and $2=\downarrow$. The gamma matrices are given in terms of the Pauli matrices, as $\gamma_0=\sigma_1$, $\gamma_5=-\sigma_3$, and $\Delta(x)$ is a (real) gap related to lattice vibrations. $\lambda_{\rm TLM} = \frac{ 2 \alpha^2}{\pi t_0 K}$ is a dimensionless coupling, where $\alpha$ is the $\pi$-electron-phonon coupling constant of the original SSH Hamiltonian, and $K$ is the elastic chain deformation constant. The equivalence between the TLM and the Gross-Neveu (GN) model~\cite{GN}, is established by setting $\lambda_{\rm TLM} = \frac{\lambda_{\rm GN}}{N \pi} $. Note that the electron-phonon interaction term in Eq.~(\ref{LagTLM}) is an analog of the fermion-boson interaction in the field theory context, which appears in different models and dimensions. In four space-time dimensions, for example, this interaction has been investigated in the framework of the linear sigma model at finite temperature~\cite{NPB}.

The GN model has been investigated earlier at finite temperature and density several times (see for instance Refs.~\cite{Wolff,Treml,Klimenko}), and recently considering also finite corrections to the leading order in the large $N$ approximation~\cite{PRB}. However, these approximations did not consider the effects of an external Zeeman magnetic field applied on the system, which is of fundamental importance in many physical situations, as in the investigation of metal-insulator transitions~\cite{Dent} and magnetization in 2D electron systems~\cite{PRB2}.

In order to consider the application of an external Zeeman magnetic field to the system and its effects, it is convenient to start by writing the grand canonical
partition function associated with ${\cal L}_{\rm GN}$,

\begin{equation}
\label{action1}
{\cal Z}= \int  ~ {\cal D} \bar \psi ~ {\cal D} \psi~ exp \left\{ \int_0^{ \beta} d\tau \int dx~ \left[ { L}_{\rm GN} \right] \right\},
\end{equation}
where $\bar \psi \equiv {\psi}^\dagger \gamma_0$, $\beta=1/k_BT$, $k_B$ is the Boltzmann constant, and $L_{\rm GN}$ is the Euclidean GN Lagrangian density:

\begin{equation}
\label{L1}
{ L}_{\rm GN}= \sum_{j=1,2} \bar \psi_j [- \gamma_0 \hbar \partial_\tau + i \hbar v_F \gamma_1 \partial_x -  \Delta(x) +  \gamma_0 \mu_j  ] \psi_j -\frac{1}{ \hbar v_F \lambda_{\rm GN}}  \Delta^2(x),
\end{equation}
where $\mu_\uparrow= \bar \mu + \delta \mu$, $\mu_\downarrow= \bar \mu - \delta \mu$. The Zeeman splitting energy is given by $\Delta E= S_z g \mu_B B_0$~\cite{Kittel}, where $S_z=\pm 1/2$, $g$ is the effective $g$-factor and $\mu_B=e \hbar/2m \approx  5.788 \times 10^{-5} ~{\rm eV}~ {\rm T}^{-1}$ is the Bohr magneton, giving $\delta \mu = \frac{g}{2} \mu_B B_0$. In \cite{NPB1} we also have chosen $\bar \mu=\mu_c$.

Integrating over the fermion fields leads to 

\begin{equation}
\label{action2}
{\cal Z} =  exp~{ \left\{ -\frac{\beta}{ \hbar v_F \lambda_{\rm GN}} \int dx ~ \Delta^2(x) \right\}} ~\Pi_{j=1}^2 det D_j ,
\end{equation}
where $D_j=- \gamma_0 \partial_\tau + i \hbar v_F \gamma_1 \partial_x +  \gamma_0 \mu_j -  \Delta(x) $ is the Dirac operator at finite temperature and density. Since $\Delta(x)$ is static, we can transform $D_j$ to the $\omega_n$ plane, where $\omega_n=(2n+1)\pi T$ are the Matsubara
frequencies for fermions, yielding $D_j= (- i \omega_n + \mu_j)\gamma_0 + i \hbar v_F \gamma_1 \partial_x  -  \Delta(x) $. After using an elementary identity $\ln (det(D_j))={\rm Tr} \ln(D_j)$, one can define the bare effective action for the static $\Delta(x)$ condensate

\begin{equation}
\label{action3}
S_{eff} [\Delta]= -\frac{\beta}{ \hbar v_F \lambda_{\rm GN}}  \int dx ~ \Delta^2(x) + \sum_{j=1}^2 {{\rm Tr} \ln (D_j)},
\end{equation}
where the trace is to be taken over both Dirac and functional indices. The condition to find the stationary points of $S_{eff} [\Delta]$ reads

\begin{equation}
\label{action4}
\frac{\delta S_{eff} [\Delta]}{\delta \Delta(x)}=0= -\frac{2 \beta}{ \hbar v_F \lambda_{\rm GN}} \Delta(x) + \frac{\delta}{\delta \Delta(x)} \left[ \sum_{j=1}^2 {{\rm Tr} \ln (D_j)} \right].
\end{equation}
The equation above is a complicated and generally unknown functional equation for $\Delta(x)$, whose solution has been investigated at various times in the literature~\cite{Dashen,Feinberg,gnpolymers,gnpolymers2,gnpolymers3,Basar}. Its solution is not only of academic interest, but has direct application in condensed matter physics as, for example, in~\cite{Review,gnpolymers3}, and in the present work.

\section{The Renormalized Effective Potential at Finite Temperature}

\subsection{Homogeneous $\Delta(x)$ Condensates}

For a constant $\Delta$ field the Dirac operator reads $D_j= (- i \omega_n + \mu_j)\gamma_0 + i \hbar v_F \gamma_1 p  -  \Delta $, so the trace in Eq.~(\ref{action3}) can be evaluated in a closed form for the asymmetrical ($\delta \mu \neq 0$) system~\cite{NPB1}. From Eq.~(\ref{action2}) one obtains the ``effective" potential $V_{eff}=-\frac{k_B T}{L} \ln {\cal Z}$, where $L$ is the length of the system:

\begin{eqnarray}
V_{eff}(\Delta,\mu_{\uparrow,\downarrow},T)=\frac{1}{ \hbar v_F \lambda_{\rm GN}} \Delta^2 
&-& k_B T \int^{+\infty}_{-\infty}{\frac{dp}{2\pi \hbar}}~ \Big[ 2 \beta E_p
+  \ln \left(1+e^{-\beta E_\uparrow^+}\right)+  \ln \left(1+e^{-\beta E_\uparrow^-}\right)\\
\nonumber
&+& \ln \left(1+e^{-\beta E_\downarrow^+}\right) + \ln \left(1+e^{-\beta E_\downarrow^-}\right) \Big],
\label{poteff}
\end{eqnarray}
where $E_{\uparrow,\downarrow}^{\pm} \equiv E_p \pm \mu_{\uparrow,\downarrow}$, $E_p = \sqrt{v_F^2 p^2+\Delta^2}$.

The first term in the integration in $p$, corresponding to the vacuum part ($\mu_{\uparrow,\downarrow}=T=0$), is divergent. Introducing a momentum cutoff $\Lambda$ to regulate this part of $V_{eff}$, we obtain, after renormalization, a finite effective potential

\begin{eqnarray}
V_{eff}(\Delta)=\frac{\Delta^2}{ \hbar v_F } \left(\frac{1}{\lambda_{\rm GN}} - \frac{3}{2 \pi} \right) + \frac{\Delta^2}{\pi \hbar v_F } \ln \left( \frac{\Delta}{m_F} \right),
\label{poteff2}
\end{eqnarray}
where $m_F$ is an arbitrary renormalization scale, with dimension of energy. The minimization of $V_{eff}(\Delta)$ with respect to $\Delta$ gives the well-known result for the non-trivial gap~\cite{GN}:

\begin{equation}
\Delta_0 = m_F e^{ 1- \frac {\pi}{\lambda_{\rm GN} }}.
\label{delta0}
\end{equation}
From this gap equation we see that with the experimentally measured $\Delta_0$ and $\alpha$, $t_0$ and $K$ which enters $\lambda_{\rm TLM}$, one sets the value of $m_F$. Equation~(\ref{poteff2}) can be expressed in a more convenient form in terms of $\Delta_0$ as

\begin{eqnarray}
V_{eff}(\Delta)=\frac{\Delta^2}{2 \pi \hbar v_F } \left[ \ln \left( \frac{\Delta^2}{\Delta_0^2} \right)-1 \right],
\label{poteff2_1}
\end{eqnarray}
which is clearly symmetric under $\Delta \to -\Delta$, which generates the discrete chiral symmetry of the GN model. As has been pointed out before~\cite{Dashen}, this discrete symmetry is dynamically broken by the non-perturbative vacuum, and thus there is a kink solution interpolating between the two degenerate minima $\Delta = \pm \Delta_0$ of (\ref{poteff2_1}) at $x = \pm \infty$:

\begin{equation}
\Delta(x)=\Delta_0 \tanh(\Delta_0 x).
\label{kink}
\end{equation}
In the next subsection we discuss the effects of space dependent $\Delta(x)$ condensates.

We can rewrite $V_{eff}(\Delta,\mu_{\uparrow,\downarrow},T)$ as

\begin{equation}
V_{eff}(\Delta,\mu_{\uparrow,\downarrow},T)= V_{eff}(\Delta) + V_{eff}(\mu_{\uparrow,\downarrow},T),
\label{poteff3}
\end{equation}
where

\begin{eqnarray}
V_{eff}(\mu_{\uparrow,\downarrow},T)=
- k_B T \int^{\infty}_{0} \frac{dp}{\pi \hbar } ~ \Big[  \ln \left(1+e^{-\beta E_\uparrow^+}\right)+ \ln \left(1+e^{-\beta E_\uparrow^-}\right)
+ \ln \left(1+e^{-\beta E_\downarrow^+}\right) + \ln \left(1+e^{-\beta E_\downarrow^-}\right) \Big].
\label{poteff4}
\end{eqnarray}
Since we can not calculate expression (\ref{poteff4}) in a closed form, we shall use a high temperature expansion to evaluate it. Using the function

\begin{equation}
I(a,b)=\int_0^{\infty} dx \left[ \ln \left(1+ e^{-\sqrt{x^2+a^2}-b} \right) + \ln \left(1+ e^{-\sqrt{x^2+a^2}+ b} \right) \right],
\label{int1}
\end{equation}
where $a=\Delta/k_BT$, and $b=\mu/k_BT$, which can be expanded in the high temperature limit, $a<<1$ and $b<<1$, yielding, up to order $a^4$ and $b^2$~\cite{Rudnei},

\begin{equation}
I(a<<1,b<<1)= \frac{\pi^2}{6}+\frac{b^2}{2}-\frac{a^2}{2}\ln \left(\frac{\pi}{a} \right) -\frac{a^2}{4}(1-\gamma_E)-\frac{7 \xi(3)}{8 \pi^2} a^2 \left(b^2+\frac{a^2}{4}  \right) + \frac{186~\xi(5)}{128 \pi^4}  b^2 a^4 + {\cal O}\left(a^2 b^4 \right),
\label{int2}
\end{equation}
where $\gamma_E \approx 0.577... $ is the Euler constant and $\xi(n)$ is the Riemann zeta function, having the values $\xi(3) \approx 1.202$, and $\xi(5) \approx 1.037$. With the equation above, together with Eq.~(\ref{poteff2}), the high temperature asymmetrical effective potential is written as

\begin{eqnarray}
V_{eff}(\Delta,\mu_{\uparrow,\downarrow},T) &\equiv& V_{eff} = \frac{\Delta^2}{\pi \hbar v_F} \left[ \ln \left( \frac{\pi k_B T}{\Delta_0} \right) - \gamma_E \right]
-\frac{\pi}{3 \hbar v_F} (k_B T)^2 \\
\nonumber
&-& \frac{1}{2 \pi \hbar v_F} \left[ \mu_{\uparrow}^2+\mu_{\downarrow}^2 
- \frac{7 \xi(3)}{8 \pi^2} \frac{\Delta^4}{(k_B T)^2} - \frac{7 \xi(3)}{4 \pi^2} (\mu_{\uparrow}^2+\mu_{\downarrow}^2) \frac{\Delta^2}{(k_B T)^2} +
\frac{186 \xi(5)}{64 \pi^4} (\mu_{\uparrow}^2+\mu_{\downarrow}^2) \frac{\Delta^4}{(k_B T)^4} \right].
\label{poteff5}
\end{eqnarray}
The equation above may be rearranged in the form of a Ginzburg-Landau (GL) expansion of the grand potential density, which is appropriate to the analysis of the phase diagram in the region near the tricritical point,

\begin{equation}
\label{Veff}
V_{eff}= \alpha_0 + \alpha_2 \Delta^2 + \alpha_4 \Delta^4,
\end{equation}
where

\begin{eqnarray}
\label{coef}
\alpha_0(\mu_{\uparrow,\downarrow},T) &=& - \frac{1}{2 \pi \hbar v_F} \left[ \mu_{\uparrow}^2+\mu_{\downarrow}^2 \right] -\frac{\pi}{3 \hbar v_F} (k_B T)^2, \\
\nonumber
\alpha_2(\mu_{\uparrow,\downarrow},T) &=& \frac{1}{\pi \hbar v_F} \left[ \ln \left( \frac{\pi k_B T}{e^{\gamma_E} \Delta_0} \right) + \frac{7 \xi(3)}{8 \pi^2} \frac{(\mu_{\uparrow}^2+\mu_{\downarrow}^2)}{(k_B T)^2} \right], \\
\alpha_4(\mu_{\uparrow,\downarrow},T) &=& - \frac{1}{32 \pi^3 \hbar v_F (k_B T)^2} \left[ -14 \xi(3) + \frac{186 \xi(5)}{4 \pi^2} \frac{(\mu_{\uparrow}^2+\mu_{\downarrow}^2)}{(k_B T)^2} \right].
\end{eqnarray}
Extremizing $V_{eff}$ we find the trivial solution ($\Delta=0$) and the chemical potential and temperature dependent gap equation

\begin{equation}
\label{poteff6}
\Delta(\mu_{\uparrow,\downarrow},T)^2= -\frac{\alpha_2}{2 \alpha_4},
\end{equation}
which has meaning only if the ratio $\frac{\alpha_2}{ \alpha_4}$ is negative. Besides, a stable configuration (i.e., bounded from below) requires, up to this order, $\alpha_4>0$. At the minimum $V_{eff}$ reads

\begin{equation}
\label{Veff2}
V_{eff, min}= \alpha_0 - \frac{\alpha_2^2}{4 \alpha_4}.
\end{equation}

\begin{figure}[htb]
  \vspace{0.5cm}
\epsfysize=5.5cm
  \epsfig{figure=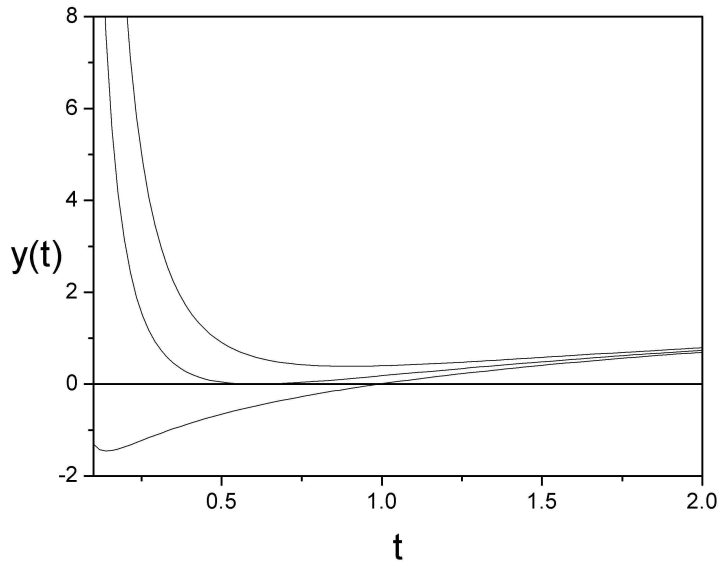,angle=0,width=12.2cm}
\caption[]{\label{y} The function $y(t)$, as a function of $t=\frac{T}{T_c(0)}$, for different values of $\eta$. The bottom curve is for $\eta=0.01$, the second curve is for $\eta=\eta_{tc}=\frac{1}{2e}\approx 0.184$, and the top curve (with no solution) is for $\eta=0.4$.}
\end{figure}

\begin{figure}[htb]
  \vspace{0.5cm}
\epsfysize=5.5cm
  \epsfig{figure=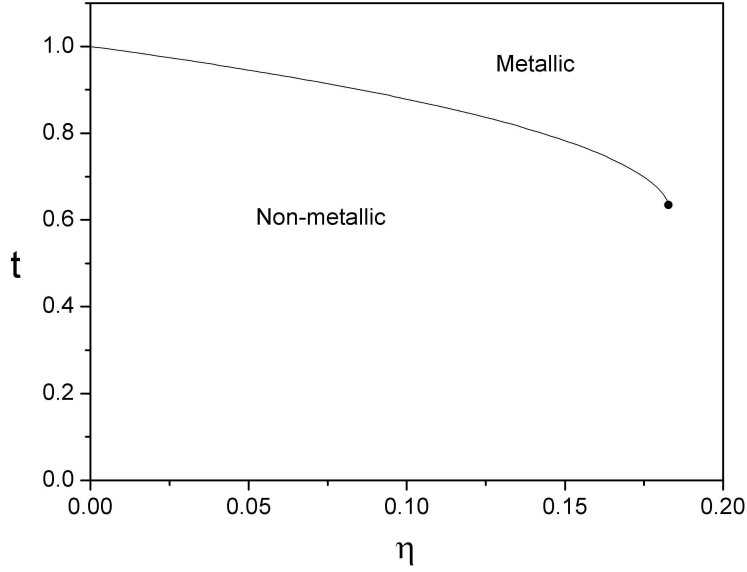,angle=0,width=12.2cm}
\caption[]{\label{pd1} The phase diagram $t=\frac{T}{T_c(0)}$ as a function of $\eta=\frac{7 \xi(3)}{8 \pi^2} \frac{(\mu_{\uparrow}^2+\mu_{\downarrow}^2)}{(k_B T_c(0))^2}$, from Eq.~(\ref{Tc1}). The small dot at the end of the second order transition line represents the tricritical point $P_{tc}=(\frac{1}{2e},\frac{1}{\sqrt{e}})$. Below this point the transition is of first order.}
\end{figure}

The critical temperature $T_c$ is, by definition, the temperature at which the gap vanishes. Thus, at $T_c$ we have $\alpha_2=0$ or

\begin{equation}
 \ln \left( \frac{\pi k_B T_c}{e^{\gamma_E} \Delta_0} \right) + \frac{7 \xi(3)}{8 \pi^2} \frac{(\mu_{\uparrow}^2+\mu_{\downarrow}^2)}{(k_B T_c)^2} =0,
\label{Tc1}
\end{equation}
where $T_c=T_c(\mu_{\uparrow},\mu_{\downarrow})$. As will become clear below, the equation above defines a second-order transition line separating the non-metallic ($\Delta \neq 0$) and metallic phases ($\Delta = 0$). At $\mu_{\uparrow}=\mu_{\downarrow}=0$, we recover the well-known result for the temperature at which the discrete chiral symmetry is restored~\cite{Jacobs}:

\begin{equation}
T_c(\mu_{\uparrow}=\mu_{\downarrow}=0) \equiv T_c(0)=\frac{e^{\gamma_E}}{\pi} \frac{\Delta_0}{k_B}.
\label{Tc2}
\end{equation}
In order to find $T_c(\mu_{\uparrow},\mu_{\downarrow})$ we define dimensionless variables $\eta=\frac{7 \xi(3)}{8 \pi^2} \frac{(\mu_{\uparrow}^2+\mu_{\downarrow}^2)}{(k_B T_c(0))^2}$ and $t=\frac{T}{T_c(0)}$, and with the help of Eq.~(\ref{Tc2}) we rewrite the L.H.S. of Eq.(\ref{Tc1}) as

\begin{equation}
y(t)=\ln(t) + \frac{\eta}{t^2}.
\label{Tc3}
\end{equation}
The zeros of $y(t)$ for a given $\eta$, i.e., for a given $\mu_{\uparrow}^2 + \mu_{\downarrow}^2$, are the respective $T_c$. This defines the (second-order) $T_c$ versus $\mu_{\uparrow}^2 + \mu_{\downarrow}^2$ phase diagram. As can be seen in Fig.~\ref{y}, the graphical analysis of $y(t)$ shows that there is no solution for this function for $\eta$ above certain value, that we define $\eta_{tc}$. Besides, at $\eta_{tc}$ we have $y=y'=0$. These two equations give $t_{tc}$ and $\eta_{tc}$ for the tricritical point $P_{tc}=(\eta_{tc},t_{tc})$. Strictly speaking, $y$ and $y'$ are associated with the coefficients of the second-order and forth-order terms of the effective potential expanded in powers of $\Delta$~\cite{jstat}. Solving the equations $y=0$ and $y'=0$ self-consistently (which is equivalent to solve $\alpha_2=\alpha_4=0$), we obtain

\begin{equation}
\eta_{tc}=\frac{7 \xi(3)}{8 \pi^2} \frac{(\mu_{\uparrow}^2 + \mu_{\downarrow}^2)_{tc}}{(k_B T_c(0))^2}=\frac{1}{2e},~~~~~~~t_{tc}=\frac{T_{tc}}{T_c(0)}=\sqrt{2\eta_{tc}}=\frac{1}{\sqrt{e}}.
\label{Tc4}
\end{equation}
For $\eta$ above certain value and less than $\eta_{tc}$, the function $y$ presents two solutions (not shown in Fig.~\ref{y}) for $T_c$. However, the lower of these always corresponds to unstable solutions. The second-order transition curve, defined as the line starting at the point $(0,T_c(0))$ and ending at the point $((\mu_{\uparrow}^2+\mu_{\downarrow}^2)_{tc},T_{tc})$, comes simply from the solution of the gap equation. This curve, shown in Fig.~\ref{pd1}, represents a system at finite temperature where the chemical potentials $\mu_{\uparrow}^2+\mu_{\downarrow}^2=2(\bar\mu^2+{\delta \mu}^2)$ start from zero and increases until $(\mu_{\uparrow}^2+\mu_{\downarrow}^2)_{tc}$. Note that $\bar\mu^2+{\delta \mu}^2$ is zero if and only if $\bar\mu^2$ and ${\delta \mu}^2$ are both zero. It is well known that below the tricritical point one has to properly minimize the effective potential rather than using the gap equation as the transition becomes first order. Thus Eq.~(\ref{Tc1}) cannot be used for finding $T_c$ below $P_{tc}$ since this equation is valid only for the second-order transition. In this case $T_c$ has to be find numerically, through the equality $V_{eff}(\mu_{\uparrow},\mu_{\downarrow},\Delta=\Delta_{min},T_c)=V_{eff}(\mu_{\uparrow},\mu_{\downarrow},\Delta=0,T_c)$, where $\Delta_{min}$ is the non-trivial minimum of $V_{eff}$.

The number densities $n_{\uparrow,\downarrow}= -\frac{\partial}{\partial \mu_{\uparrow,\downarrow}} V_{eff}(\Delta,\mu_{\uparrow,\downarrow},T)$ read

\begin{equation}
n_{\uparrow,\downarrow}= \int^{\infty}_{0}{\frac{dp}{\pi \hbar}}~ \left[ n_k(E_{\uparrow,\downarrow}^{-}) - n_k(E_{\uparrow,\downarrow}^{+}) \right],
\label{n1}
\end{equation}
where $n_k(E_{\uparrow,\downarrow}^{+,-})=\frac{1}{e^{\beta E_{\uparrow,\downarrow}^{+,-}}+1}$ is the Fermi distribution function. The density difference

\begin{equation}
\delta n = n_{\uparrow}-n_{\downarrow}
\label{n2}
\end{equation}
is zero if $\delta \mu = \frac{g}{2} \mu_B B_0=0$, at any temperature, since in this case we have the equalities $n_k(E_{\uparrow}^{+})=n_k(E_{\downarrow}^{+})$, and $n_k(E_{\uparrow}^{-})=n_k(E_{\downarrow}^{-})$. The physical meaning of these results is that at zero external Zeeman magnetic field, the $\uparrow$ (up) and $\downarrow$ (down) electrons of the conduction $(+)$ band have the same density, and the same for the electrons of the valence $(-)$ band.

In the high temperature regime, the number densities are given by 

\begin{equation}
n_{\uparrow,\downarrow}(T)= \frac{1}{ \pi \hbar v_F} \left[ 1 - \frac{7 \xi(3)}{4 \pi^2} \frac{\Delta^2}{(k_B T)^2} \right]\mu_{\uparrow,\downarrow}.
\label{nd1}
\end{equation}
In the high temperature limit the total number density, $n_{T}(T)=n_{\uparrow}(T)+n_{\downarrow}(T)$, is independent of the applied field

\begin{equation}
n_{T}(T)= \frac{2}{ \pi \hbar v_F} \left[ 1 - \frac{7 \xi(3)}{4 \pi^2} \frac{\Delta^2}{(k_B T)^2} \right]\bar \mu,
\label{nT}
\end{equation}
and for the density difference we obtain

\begin{equation}
\delta n_{high~T}(T, \delta \mu)= \frac{2}{ \pi \hbar v_F} \left[ 1 - \frac{7 \xi(3)}{4 \pi^2} \frac{\Delta^2}{(k_B T)^2} \right]\delta \mu,
\label{nd2}
\end{equation}
that, as we have observed before, is clearly zero if $B_0=\delta \mu = 0$. Since the densities have to be evaluated at the minimum of the effective potential, we use Eq.~(\ref{poteff6}) in the equation above and find the temperatures at which the densities and, consequently, the density difference vanish. These temperatures are the solutions of

\begin{equation}
\gamma_E -\frac{1}{2} + \ln \left( \frac{\Delta_0}{\pi k_B T_c^*} \right) - \frac{7 \xi(3)}{8 \pi^2} \frac{(\mu_{\uparrow}^2+\mu_{\downarrow}^2)}{(k_B T_c^*)^2} =0,
\label{nd3}
\end{equation}
where $T_c^*=T_c^*(\mu_{\uparrow},\mu_{\downarrow})$. As for the gap parameter, the equation above defines the second-order line for the densities and the density imbalance. At $\mu_{\uparrow}=\mu_{\downarrow}=0$, we get

\begin{equation}
T_c^*(\mu_{\uparrow}=\mu_{\downarrow}=0) \equiv T_c^*(0)=\frac{e^{\gamma_E-\frac{1}{2}}}{\pi} \frac{\Delta_0}{k_B}.
\label{Tc*1}
\end{equation}
It is very easy to see that $T_c^*(0)=\frac{T_c(0)}{\sqrt{e}}$, which coincides with $T_{tc}$, where $T_{tc}$ is given by Eq.~(\ref{Tc4}). Proceeding as before we find

\begin{equation}
\eta_{tc}^*=\frac{7 \xi(3)}{8 \pi^2} \frac{(\mu_{\uparrow}^2+\mu_{\downarrow}^2)_{tc}}{(k_B T_c^*(0))^2}=\frac{1}{2e^2},~~~~~~~t_{tc}^*=\frac{T_{tc}^*}{T_c^*(0)}=\sqrt{2\eta_{tc}^*}=\frac{1}{e},
\label{nd4}
\end{equation}
defining the tricritical point for the densities, total density and density imbalance second order curves.

Let us now verify the possibility of a fully polarized state at finite temperature. It would be possible with a magnetic field with a intensity such that $n_{\downarrow}$ in Eq.~(\ref{nd1}) vanishes. In this case $\mu_{\downarrow}=\bar \mu-\delta \mu_c=\mu_c- \frac{g}{2} \mu_B B_{0,c}=0$, or

\begin{equation}
B_{0,c}= \frac{2 \mu_c}{g \mu_B},
\label{cmf}
\end{equation}
yielding, for TPA (for which $\mu_c=\frac{\Delta_0}{\sqrt{2}}$ and $g \approx 2$) a critical magnetic field

\begin{equation}
B_{0,c} \approx 8.6 ~{\rm k T},
\label{cmf2}
\end{equation}
which is, as the critical magnetic field found in \cite{NPB1} at zero temperature ($\approx 4.6 ~{\rm k T}$), a magnetic field of very high magnitude, compared to the maximum current laboratory values~\cite{lab}.

\subsection{Inhomogeneous $\Delta(x)$ Condensates}

Since we consider the addition of a chemical potential (i.e., doping) in the theory representing Peierls distorted Q1DM, and the effects of a Zeeman magnetic field on these materials, some important remarks are in order. It is well-known that doping in conducting polymers with degenerate ground states results in lattice deformation, or non-linear excitations, such as kink solitons and polarons, meaning that $\Delta(x)$ can vary in space~\cite{Horovitz1,Horovitz2,Review}. Therefore, one may expect not only homogeneous-like configurations (as considered in the previous subsection), but also that the inclusion of these excitations in any theoretical calculation in this model should be considered. In this context, within the GN field theory model that we are considering, by taking into account kink-like configurations in the large $N$ approximation, the authors of Refs.~\cite{gnpolymers,gnpolymers2,gnpolymers3,Basar} found evidence for a crystalline phase that shows up in the extreme $T \sim 0$ and large $\mu$ part of the phase diagram, while the other extreme of the phase diagram, for large $T$ and small $\mu$, seemed to remain identical to the usual large $N$ results for the critical temperature and tricritical points, which are well-known results \cite{Wolff} for the GN model. 

To take into account the effects of inhomogeneous configurations in the GL expansion of the grand potential density, let us write Eq.~(\ref{Veff}) in terms of $\Delta(x)$ and its derivatives up to $\alpha_4$~\cite{gnpolymers,gnpolymers2,gnpolymers3,Basar}:

\begin{equation}
\label{Veff(x)}
V_{eff}(x)= \alpha_0 + \alpha_2 \Delta(x)^2 + \alpha_4 [ \Delta(x)^4 + \Delta(x)'^2 ],
\end{equation}
where $\Delta(x)' \equiv d \Delta(x) / dx$. A straightforward variational calculation gives the following condition for the minimization of the free energy $E = \int V_{eff}(x)~dx$:

\begin{equation}
\label{difeq1}
\Delta(x)'' -2 \Delta(x)^3 - \frac{\alpha_2}{\alpha_4}\Delta(x)=0.
\end{equation}
The general solution of an equation of the form

\begin{equation}
\label{difeq2}
\Delta(x)'' -2 \Delta(x)^3 + (1+\nu) \Delta_0^2 \Delta(x)=0,
\end{equation}
can be written as~\cite{Basar}

\begin{equation}
\label{difeq3}
\Delta(x)= \Delta_0 \sqrt{\nu} {\rm sn} (\Delta_0 x; \nu),
\end{equation}
where ${\rm sn}$ is the Jacobi elliptic function with the real elliptic parameter $0 \leq \nu \leq  1$. The ${\rm sn}$ function has period $2 {\bf K}(\nu)$, where ${\bf K}(\nu) \equiv \int_0^{\pi/2}[1-\nu \sin^2(t)]^{-1/2} dt$ is the complete elliptic integral of first kind. $\Delta(x)$ in (\ref{difeq3}) represents an array of real kinks. When $\nu=1$ Eq.~(\ref{difeq3}) is reduced to the single kink condensate given in Eq.~(\ref{kink}). By comparing Eqs.~(\ref{difeq1}) and (\ref{difeq2}) one can identify the scale parameter $\Delta_0$ as

\begin{equation}
\label{difeq4}
\Delta_0^2=\left( -\frac{\alpha_2}{\alpha_4} \right) \left( \frac{1}{1+\nu} \right).
\end{equation}
Given that $\frac{1}{1+\nu} > 0$, the solution for inhomogeneous condensates has physical meaning only if the ratio $\frac{\alpha_2}{\alpha_4}$ is negative, as in the case of homogeneous condensates. In terms of Eq.~(\ref{difeq3}) it can be shown that the $x$ dependent grand potential density can be written as

\begin{equation}
\label{Veff(x)2}
V_{eff}(x)= \alpha_0 + \alpha_2\Delta(x)^2 + \alpha_4 \frac{1}{3} \left[ (1+\nu)\Delta_0^2 \Delta(x)^2 + \nu \Delta_0^4 \right].
\end{equation}
Averaging over one period, it is found~\cite{Basar} that $<\Delta(x)^2>= \left(1-\frac{{\bf E}(\nu)}{{\bf K}(\nu)} \right) \Delta_0^2 $, where ${\bf E}(\nu)$ is the complete elliptic integral of second kind. The ratio ${\bf E}(\nu)/{\bf K}(\nu)$ is a smooth function of $\nu$ interpolating monotonically between $0$ and $1$. Thus we can write

\begin{equation}
\label{Veff(x)3}
<V_{eff}(x)> = \alpha_0 +  {\cal A}_2 \Delta_0^2  + {\cal A}_4 \Delta_0^4,
\end{equation}
where 

\begin{eqnarray}
\label{Newcoef}
{\cal A}_2 &=& \alpha_2 \left(1-\frac{{\bf E}(\nu)}{{\bf K}(\nu)} \right),\\
\nonumber
{\cal A}_4 &=& \alpha_4 \frac{1}{3} \left[\nu + (1+ \nu)\left(1-\frac{{\bf E}(\nu)}{{\bf K}(\nu)} \right)\right].
\end{eqnarray}
\newline
The interesting results obtained considering inhomogeneous condensates are:
\newline
\newline 
(${\bf 1.}$) For $\nu=1$, $\frac{{\bf E}(\nu=1)}{{\bf K}(\nu=1)}=0$, so the grand potential density is that of the homogeneous case, Eq.~(\ref{Veff}), at the non-trivial minimum.
\newline
(${\bf 2.}$) For $\nu=0$, $\frac{{\bf E}(\nu=0)}{{\bf K}(\nu=0)}=1$, so the grand potential density is that of the metallic phase, for which $\Delta=0$ and $V_{eff}=\alpha_0$.
\newline
(${\bf 3.}$) The tricritical point is still found for $\alpha_2=\alpha_4=0$. These coefficients are $\mu_{\uparrow,\downarrow}$ and $T$ dependent and were not affected by the space dependence of the condensate $\Delta(x)$. Then the location of the tricritical point in a Peierls distorted Q1DM under the influence of an external Zeeman magnetic field is unaltered even considering a $x$ dependent grand potential density. This happens because in the high temperature limit the influence of the Zeeman field is not sufficient to change the position of the tricritical point. This same conclusion has been obtained for the symmetric case ($\delta \mu =B_0=0$)~\cite{gnpolymers,gnpolymers2,gnpolymers3,Basar}.

\subsection{Magnetic Properties}

The Pauli magnetization of the chain in the high temperature limit has the following expression:

\begin{eqnarray}
M_{high~T}(T) &=& \mu_B \delta n_{high~T}(T)= \frac{2 \mu_B}{ \pi \hbar v_F} \left[ 1 - \frac{7 \xi(3)}{4 \pi^2} \frac{\Delta^2}{(k_B T)^2} \right]\delta \mu \nonumber \\
&=& \frac{2 g \mu_B^2}{ \pi \hbar v_F} \left[\ln \left( \frac{T}{T_c^*(0)} \right) + \frac{7 \xi(3)}{8 \pi^2} \frac{(\mu_{\uparrow}^2+\mu_{\downarrow}^2)}{(k_B T)^2} \right] B_0,
\label{mag1}
\end{eqnarray}
where we have made use of Eq.~(\ref{poteff6}) to leading order in $\frac{(\mu_{\uparrow}^2+\mu_{\downarrow}^2)}{(k_B T)^2}$. The second-order line where the magnetization vanishes is the same as the one given by Eq.~(\ref{nd3}). Finally, we obtain the magnetic susceptibility in this regime

\begin{equation}
\chi_{high~T}(T)= \frac{\partial M_{high~T}(T)}{\partial B_0}= \chi(0)+ \chi(T),
\label{ms1}
\end{equation}
where

\begin{equation}
\chi(0)=\frac{g \mu_B^2}{ \pi \hbar v_F},
\label{ms2}
\end{equation}
and

\begin{equation}
\chi(T)=\frac{2 g \mu_B^2}{ \pi \hbar v_F} \left[\ln \left(\frac{T}{T_c(0)} \right) + \frac{7 \xi(3)}{4 \pi^2(k_B T)^2} \left(\bar \mu^2 + \frac{3}{4}g^2 \mu_B^2 B_0^2 \right) \right].
\label{ms3}
\end{equation}
$\chi(0)$ is the well known zero temperature contribution for the Pauli expression of the magnetic susceptibility for noninteracting electrons. The function $\chi_{high~T}(T)$ also behaves at finite temperature as the densities and the magnetization, with a second-order transition up to a tricritical point given by Eq.~(\ref{nd4}). Below this point the transition is again of first order. 

Note that in spite of the fully polarization, at $B_{0,c}$ the magnetization is given by exactly the same expression shown in Eq.~(\ref{mag1}). With the help of Eq.~(\ref{nd1}) we find:

\begin{eqnarray}
M_{high~T,c}(T) = \mu_B n_{\uparrow}(T) &=&  \frac{\mu_B}{ \pi \hbar v_F} \left[ 1 - \frac{7 \xi(3)}{4 \pi^2} \frac{\Delta^2}{(k_B T)^2} \right](\mu_c + \delta \mu_c) \nonumber \\
&=& \frac{2 g \mu_B^2}{ \pi \hbar v_F} \left[ \ln \left( \frac{T}{T_c^*(0)} \right) + \frac{7 \xi(3)}{8 \pi^2} \frac{(\mu_{\uparrow}^2+\mu_{\downarrow}^2)}{(k_B T)^2} \right] B_{0,c}.
\label{mag2}
\end{eqnarray}
This shows that this function is indeed continuous for $0\leq B_0 \leq B_{0,c}$.

\section{Summary}

We have investigated the mean-field finite temperature phase diagram of Q1DM under the influence of an external Zeeman magnetic field. We found that the gap parameter and the magnetization (as well as the densities and density imbalance, and the magnetic susceptibility) of asymmetrically doped Q1DM have a similar second-order behavior until their respective tricritical points are reached. Below these points the transitions are of first order. We found these two tricritical points analytically. He have shown that the location of the tricritical point in the $t$ $\left(=\frac{T}{T_c(0)}\right)$ versus $\eta$ $\left(=\frac{7 \xi(3)}{8 \pi^2} \frac{(\mu_{\uparrow}^2+\mu_{\downarrow}^2)}{(k_B T_c(0))^2}\right)$ phase diagram stays at the same place by considering both homogeneous and inhomogeneous condensates, as occur in symmetric $\delta \mu = B_0 = 0$ systems~\cite{gnpolymers,gnpolymers2,gnpolymers3,Basar}. We have shown that for the particular case of TPA, in order to have a fully polarized organic conductor at finite temperature it would be necessary to have a very high critical magnetic field, namely $B_{0,c}$. However, for a given magnetic field below $B_{0,c}$, partial polarizations (magnetizations) can be realized experimentally, provided the temperatures are kept outside the ``$non-metallic$" region of Fig.~\ref{pd1}. It is worth noting that, according to Eq.~(\ref{cmf}), for other 1D systems with a smaller critical chemical potential or with a greater effective $g$-factor, a smaller (attainable) critical magnetic field necessary for a fully polarization of the Q1DM would be found. As a final remark, it would also be very interesting to study the transport properties of the asymmetrically doped Peierls distorted Q1DM at zero and finite temperature, employing the field theory approach. We intent to address these topics elsewhere.

\section{Acknowledgments}

The author acknowledges partial support by the Brazilian funding agencies CNPq and FAPEMIG. I am grateful to A. L. Mota and R. O. Ramos for stimulating conversations.

\end{document}